\documentclass[12pt,preprint]{aastex}

\usepackage{txfonts}
\usepackage{longtable}
\usepackage{rotating}
\usepackage{natbib}
\usepackage{graphicx}
\usepackage{graphics}
\usepackage{psfrag}
\usepackage{amssymb}
\bibpunct{(}{)}{;}{a}{}{,}
\def\Teff{$T_{\mathrm{eff}}$}

\def\ms{$\mathrm{m\,s}^{-1}$}

\def\M{\ensuremath{M_{\odot}}}
\def\R{\ensuremath{R_{\odot}}}
\def\Rearth{\ensuremath{R_{\oplus}}}

\shorttitle{Earth-like planets in the habitable zone of cool white dwarfs}
\shortauthors{Fossati et al.}

\begin{document}


\title{The habitability and detection of Earth-like planets orbiting cool
white dwarfs.}
\author{L. Fossati\altaffilmark{1}}
\affil{Department of Physical Sciences, The Open University,
	Walton Hall, Milton Keynes MK7 6AA, UK}
\email{l.fossati@open.ac.uk}
\and
\author{S. Bagnulo}
\affil{Armagh Observatory, College Hill, Armagh BT61 9DG,
	Northern Ireland, UK}
\email{sba@arm.ac.uk}
\and
\author{C.~A. Haswell}
\affil{Department of Physical Sciences, The Open University,
	Walton Hall, Milton Keynes MK7 6AA, UK}
\email{C.A.Haswell@open.ac.uk}
\and
\author{M.~R. Patel}
\affil{Department of Physical Sciences, The Open University,
	Walton Hall, Milton Keynes MK7 6AA, UK}
\email{M.R.Patel@open.ac.uk}
\and
\author{R. Busuttil}
\affil{Department of Physical Sciences, The Open University,
	Walton Hall, Milton Keynes MK7 6AA, UK}
\email{r.busuttil@open.ac.uk}
\and
\author{P.~M. Kowalski}
\affil{GFZ German Research Centre for Geosciences, Telegrafenberg,
	14473 Potsdam, Germany}
\email{kowalski@gfz-potsdam.de}
\and
\author{D.~V. Shulyak}
\affil{Institute of Astrophysics, Georg-August-University,
	Friedrich-Hund-Platz 1, D-37077, G\"ottingen, Germany}
\email{denis.shulyak@gmail.com}
\and
\author{M.~F. Sterzik}
\affil{European Southern Observatory, Casilla 19001, Santiago 19, Chile}
\email{msterzik@eso.org}

\altaffiltext{1}{Argelander-Institut f\"ur Astronomie der Universit\"at Bonn,
Auf dem H\"ugel 71, 53121 Bonn, Germany}


%
\begin{abstract}
Since there are several ways planets can survive the giant phase of the host 
star, we examine the habitability and detection of planets orbiting white 
dwarfs. As a white dwarf cools from 6000\,K to 4000\,K, a planet orbiting 
at 0.01\,AU  would remain in the Continuous Habitable Zone (CHZ) for 
$\sim$8\,Gyr. We show that photosynthetic processes can be sustained on such 
planets. The DNA-weighted UV radiation dose for an Earth-like planet in the 
CHZ is less than the maxima encountered on Earth, hence non-magnetic white 
dwarfs are compatible with the persistence of complex life. Polarisation due 
to a terrestrial planet in the CHZ of a cool white dwarf is 10$^2$ (10$^4$) 
times larger than it would be in the habitable zone of a typical M-dwarf 
(Sun-like star). Polarimetry is thus a viable way to detect close-in rocky 
planets around white dwarfs.
Multi-band polarimetry would also allow reveal the presence of a planet 
atmosphere, providing a first characterisation. Planets in the CHZ of a 
0.6\,\M\ white dwarf will be distorted by Roche geometry, and a Kepler-11d 
analogue would overfill its Roche lobe. With current facilities a 
Super-Earth-sized atmosphereless planet is detectable with polarimetry around 
the brightest known cool white dwarf. Planned future facilities render smaller 
planets detectable, in particular by increasing the instrumental sensitivity 
in the blue.
\end{abstract}
%


%
\section{Introduction}
\label{sec:intro}
The search for habitable Earth-like planets is a major contemporary goal of 
astronomy. As the detection of exoplanets is biased towards systems with small 
differences in mass, radius and luminosity between star and planet 
\citep[e.g.,][]{Haswell10}, M-type main sequence stars have become prime 
targets in the search of Earth-like planets in the habitable zone. M dwarfs 
evolve slowly: their planets might remain within a continuously habitable 
zone (CHZ)\footnote{Range of planet orbital distances at which the
planet is habitable for a minimum of 3\,Gyr.}, i.e. harbouring surface liquid
water, for several Gyr, providing ample time for the advent of life on
a rocky planet.

With an effective temperature (\Teff) $\leq$6000\,K, cool white dwarfs (CWD)
are also promising hosts of rocky planets in the habitable zone. White dwarfs
initially cool down rapidly, with temperature decreasing by thousands of
degrees in $\sim$3\,Gyr \citep{salaris10}. At \Teff$\sim$6000\,K,
crystallisation slows the cooling process. This produces a habitable zone
which endures for up to 8\,Gyr \citep{agol}, well in
excess of the time required for life to arise on Earth. White dwarfs provide
a stable luminosity source without the potentially damaging radiation
produced by stellar activity in M dwarfs. A planet orbiting close to a white
dwarf would synchronize within 1000\,yr and would have a stable orbit, as the
planet would not raise tides on the star \citep{agol}.

The major issue for the presence of a planet close to a white dwarf is
survival during the host star's giant phase. \citet{faedi11} review several
mechanisms which would result in a planet orbiting a white dwarf.
\citet{charpinet11} found two Earth-sized bodies in a very close orbit
($\sim$0.007\,AU) around a post-red-giant star proving that planet-sized
objects can survive the post-main sequence evolution phases of their host
star. Further evidence for the existence of rocky bodies close to
white dwarfs comes from the presence of metallic lines
(e.g. Mg and Fe) in the spectra of DZ white dwarfs \citep{zuckerman}. Heavy
metals in the atmospheres of these stars can only be explained by atmospheric
``pollution" caused by the accretion of terrestrial-like planets
or planetesimals \citep[see e.g.,][]{farihi10,melis11,klein11,gensicke}. 

The low luminosity of CWDs creates a habitable zone at only $\sim$0.01\,AU,
ten times closer than for M dwarfs. This facilitates the detection of small
bodies orbiting white dwarfs: \citet{agol} showed that transits of Mars-sized
planets in the white dwarf CHZ would be 1\% deep, easily detectable with
present-day ground-based facilities, even for rather faint stars, though
searches for planetary companions to white dwarfs have so far been unsuccessful 
\citep{friedrich,faedi11}.

Polarimetric techniques can also be used to discover and characterize
exoplanets. As shown by \citet{seager00} and
\citet{stam08}, as a planet orbits around the host star, the amount of
polarisation varies regularly, showing maxima when the planet is near
quadrature. The amplitude of the variation depends mainly on the orbital
inclination, $i$, with no polarimetric variability for a face-on
orbit. The detection and measurement of regular polarisation variations permit
discovery of exoplanets, with an efficiency dependent on
$i$, similar to that of the radial velocity planet detections.
Spectropolarimetry of planet-hosting stars could characterise the atmosphere 
of an exoplanet \citep{stam08}, something now only possible for transiting 
exoplanets.
\section{Is life possible on Earth-like planets in the white dwarf CHZ?}
\label{sec:life}
It is important to establish whether CWD have a flux distribution which
would allow the advent of complex life on Earth-like planets orbiting in the
CHZ. Consider an Earth-like planet with surface liquid water and a
nitrogen-dominated atmosphere\footnote{Nitrogen would be a major
constituent also of a lifeless Earth \citep{kasting93}.}, orbiting in the white
dwarf CHZ. Major requirements for the persistence of life, as we
know it on Earth, are sufficient suitable radiation to sustain photosynthesis
and a UV flux low enough to prevent fatal DNA damage.

\begin{figure}
\includegraphics[width=16cm]{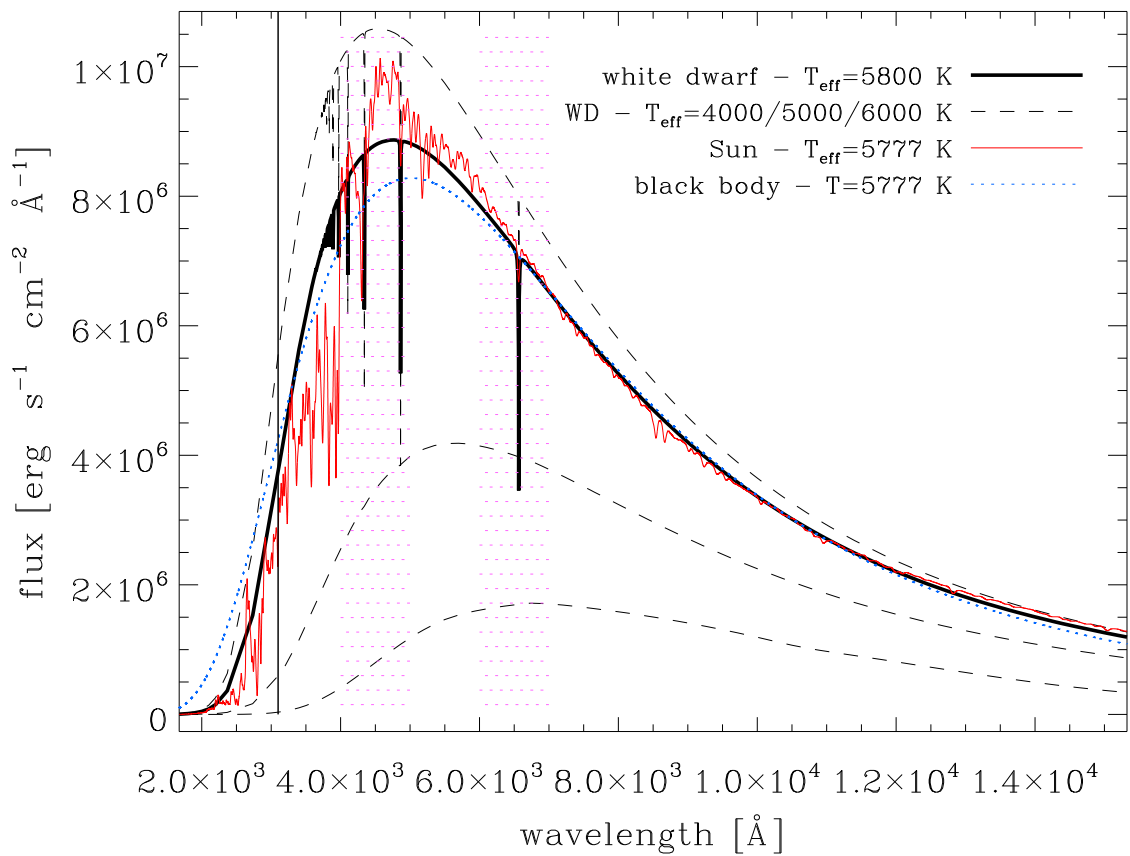}
\caption{\label{fig:fluxes} Comparison between white dwarf and solar synthetic
emergent fluxes per unit area at the emitting photosphere. The thick
line: 5800\,K hydrogen white dwarf; dashed lines: hydrogen white dwarfs
with \Teff = 4000, 5000, and 6000\,K. With decreasing temperature the hydrogen
lines weaken untill they almost disappear. Thin continuous red line: synthetic
fluxes of the Sun calculated with MARCS models; dotted line: a 5777\,K, black
body, i.e. with the Sun's \Teff. The vertical black line at 3100\,\AA\ shows
the limit for the DNA damaging fluxes. The shaded areas show the wavelengths
playing a major role in the process of photosynthesis.}
\end{figure}

Figure~\ref{fig:fluxes} shows a comparison between the synthetic surface
fluxes of CWDs with pure hydrogen atmospheres at four different \Teff, the
Sun \citep[from][]{marcs}, and a 5777\,K black body. We used CO white dwarfs
of radius 0.013\,\R, and mass of 0.6\,\M. The model fluxes were computed
by the stellar atmosphere code developed by \citet{K06} and \citet{KS06}.
%

The wavelength ranges playing a role in the photosynthetic processes
\citep[e.g.,][]{mccree} are bluewards of 5000\,\AA\ and between 6000 and
7000\,\AA. The fluxes integrated over these wavelengths for a 5800\,K white
dwarf and the Sun are almost identical. This direct comparison is valid
because the angular diameter of a CWD, as seen from the CHZ, is similar to
that of the Sun, as seen from Earth \citep{agol}. \citet{raven07} concluded
that photosynthesis can occur on exoplanets in the habitable zone of M dwarfs.
The photosynthetic relevant flux intercepted by a planet in the CHZ of a
4000\,K white dwarf is larger than that intercepted by the same planet
in the habitable zone of a typical M-dwarf. These comparisons show that
photosynthesis would be both feasible and efficient on planets orbiting
in the white dwarf CHZ.

\begin{figure}
\includegraphics[width=16cm]{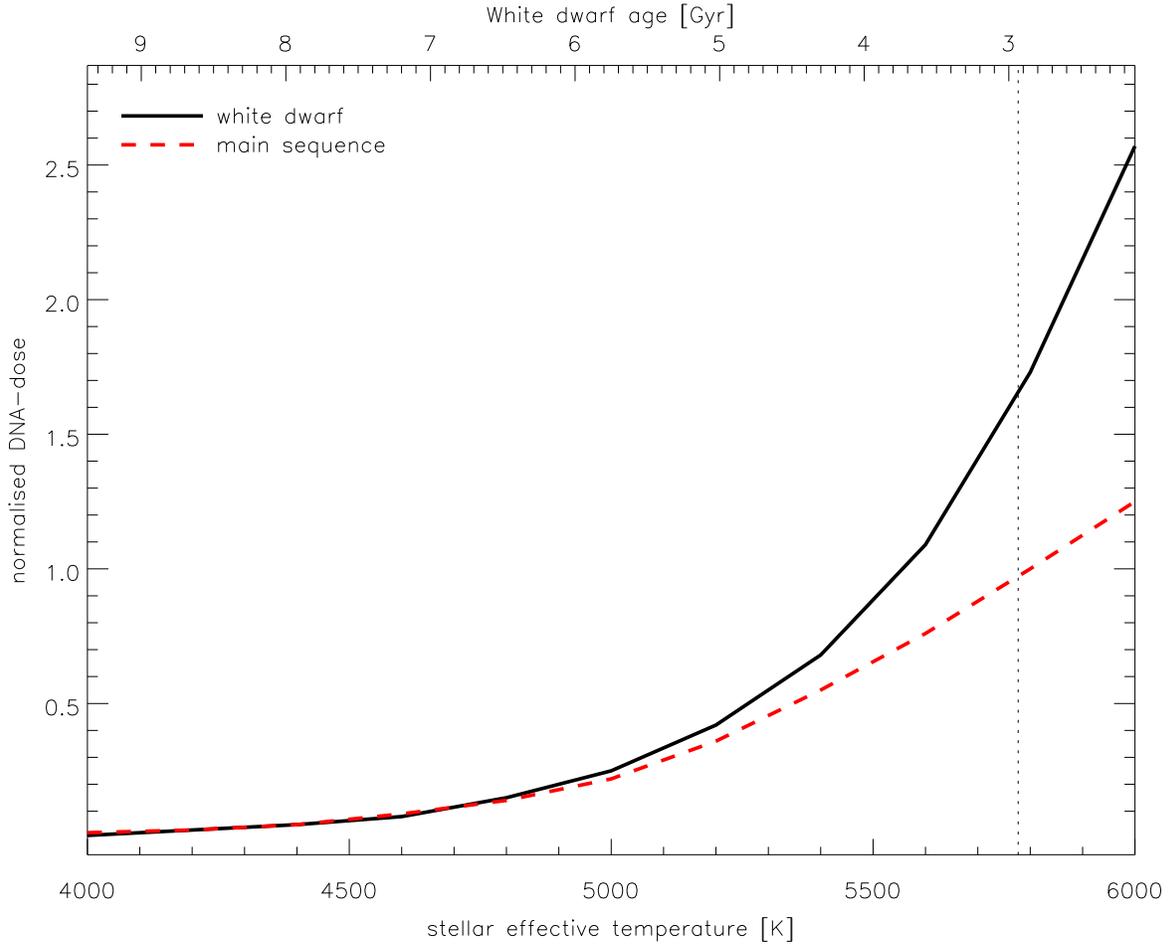}
\caption{\label{fig:dna-dose} DNA-weighted dose for an Earth-like planet
orbiting in the habitable zone of a CWD (black continuous line)
and of main sequence stars (dashed red line). In the white dwarf case we
adopted an orbital separation of 0.01\,AU, while in the main sequence case we
adopted a varying orbital separation to maintain constant the planet
equilibrium temperature. The DNA-weighted dose is normalised to that of
present Earth. The upper x-axis indicates the white dwarf age, at a given
temperature.}
\end{figure}

Fluxes shortwards of $\sim$3100\,\AA\  (see Fig.~\ref{fig:fluxes})
can potentially damage DNA molecules. To check whether the UV radiation
emitted by a CWD is a threat to the formation and persistence of DNA
molecules, we compared the DNA dose expected for an Earth-like planet in
the white dwarf CHZ with that experienced on Earth. We adapted a radiative 
transfer model developed for application to Mars \citep{patel04} to create a 
simplified Earth-atmosphere model and determine the amount of UV radiation 
received at the planet's surface as a function of white dwarf temperature. 
To provide an effective comparison as a function of stellar temperature, we 
modelled the same conditions for a hypothetical Earth-like planet around a main
sequence star, with the semi-major axis varied as a function of temperature,
keeping the ``main sequence exoplanet" within the habitable zone. This
comparison is shown in Fig.~\ref{fig:dna-dose}. We used a standard terrestrial
atmospheric composition, with a nominal ozone abundance of 300\,DU. This is
a first-order approximation as we neglect important factors such as clouds. 
To interpret the surface irradiances in a biological context, we applied a 
DNA action spectrum \citep{dna1,dna2} to the modelled spectra to
define a DNA-weighted dose for each case. DNA is used here as a generic model
of the response of carbon-based organisms to UV.
%

The DNA-weighted UV dose encountered at the surface of an Earth-like planet
in the white dwarf CHZ becomes comparable to that of an exoplanet in the
habitable zone of a main sequence star at approximately 5000\,K.
Interestingly, present-day solar conditions produce an average dose on Earth
a factor of only 1.65 less than that for a white dwarf with solar \Teff.
Varying terrestrial atmospheric conditions at times produce DNA-weighted
doses on Earth as high as that on a CWD planet. Figure~\ref{fig:dna-dose}
also clearly demonstrates that the DNA-weighted dose for a hypothetical
Earth-like planet around a CWD is remarkably benign from
an astrobiological perspective, for an extremely long period of time.

We therefore conclude that a CWD is a plausible source of energy
for the advent and persistence of complex life, as we know it on Earth.
We have to point out that about 10\% of all white dwarfs host magnetic
fields \citep{putney,landstreet2012}. A planet in the
CHZ of a magnetic white dwarf would constantly orbit inside the star's
magnetic field, which, at 0.01\,AU can be as large as a few kG. Such a field,
interacting with biological matter, might prevent the persistence of life.
This issue is the subject of a forthcoming study and for now we assume the
potential planet-hosting white dwarf to be non- (or weakly) magnetic.
This assumption also implies the absence of magnetic drag, which would
cause the planet orbit to decay.

In the previous analysis, we also assumed an atmosphere-bearing planet, 
although the initial high temperature of the white dwarf would cause the 
atmosphere of a close-in planet to evaporate. Various mechanisms, such as 
planet migration and crustal outgassing, could however replenish an atmosphere.
\section{Polarisation from an Earth-like planet in the white dwarf CHZ}
\label{sec:pol}
Using white dwarf model fluxes from Sect.~\ref{sec:life} and models of
reflected light and degree of polarisation from \citet{stam08}, we
calculated the amount of polarised light reflected by the atmosphere of an 
unresolved Earth-sized planet in the CHZ of CWDs.
Throughout we adopt a phase angle $\alpha=68^\circ$ (angle between star and 
Earth as seen from the planet) and a cloudfree Lambertian planet 
surface with a wavelength independent albedo of 1.0 
\citep[labelled as ``L10" by][]{stam08}. This is an idealised case: for 
comparison the Earth's albedo is $\approx0.3$ \citep{Charlson05}, though a 
snowball Earth might have an albedo of $\approx0.85$.
Assuming an orbit inclination angle of 90$^\circ$, the polarisation as a 
function of wavelength ($\lambda$) and phase angle, 
$P_{\mathrm{ obs}}(\lambda,\alpha)$, seen by a distant observer is:
\begin{equation}
\label{eq:pol-reduced1}
P_{\mathrm{ obs}}(\lambda,\alpha)
=
\frac{F_*\frac{\pi r^2}{4\pi d^2}b_1(\lambda,\alpha)}
{F_*(1+\frac{\pi r^2}{4\pi d^2}a_1(\lambda,\alpha))}\,\,,
\end{equation}
where $a_1(\lambda,\alpha)$ and $b_1(\lambda,\alpha)$, varying 
between 0 and 1.2, are two elements of the planet scattering matrix 
\citep[e.g.,][]{stam06}, $r$ and $d$ are respectively the planet radius and 
orbital separation, and $F_*$ is the stellar flux. The numerator represents 
the polarised flux reflected by the planet, while the denominator corresponds 
to the total flux emitted by the system 
($\approx F_*$ as $\frac{r^2}{d^2}<<1$). 

Following \citet{stam08}, $b_1(\lambda,\alpha)$ is maximum at 
$\alpha=68^\circ$, therefore the maximum polarisation becomes:
\begin{equation}
\label{eq:pol-reduced2}
P_{\mathrm{ obs,max}}(\lambda)
\approx
\frac{r^2}{d^2}\frac{1}{4}b_1(\lambda,\alpha=68^\circ)\,\,.
\end{equation}

\begin{figure}
\includegraphics[width=16cm]{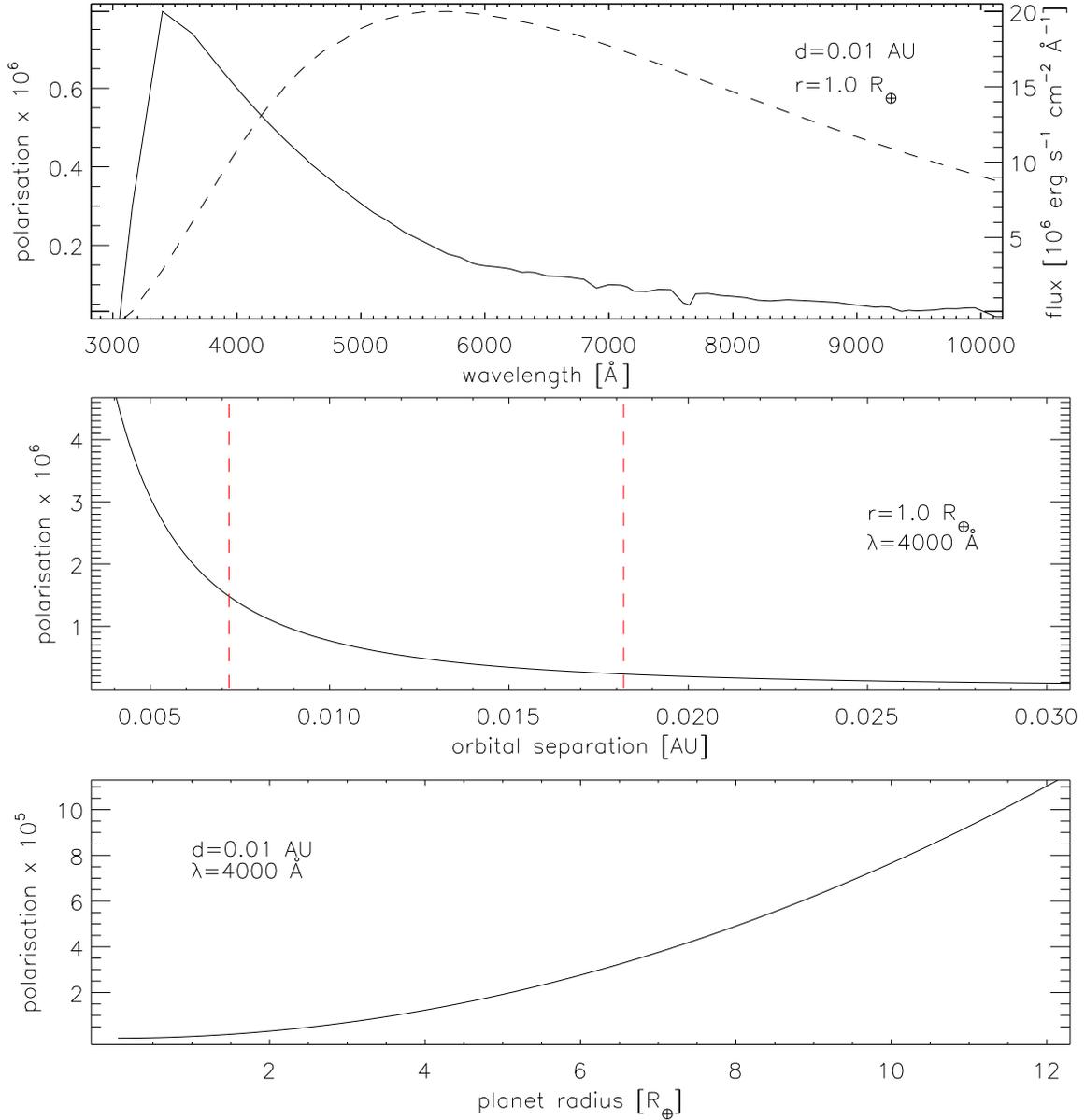}
\caption{\label{fig:polarisation} Top panel: polarisation in the
visible region as a function of wavelength for an Earth-sized planet at a
distance of 0.01\,AU from a 5000\,K white dwarf. The white dwarf synthetic
flux is shown by a dashed line (scale on the right-hand side).
Middle panel: polarisation at 4000\,\AA\ from an Earth-sized planet
as a function of distance from a 5000\,K white dwarf. The dashed vertical
lines indicate the limits of the habitable zone, calculated following
\citet{selsis07}, using the Venus and early-Mars criteria. Bottom panel:
polarisation at 4000\,\AA\ as a function of planet radius placed at a
distance of 0.01\,AU from a 5000\,K white dwarf.}
\end{figure}

Figure~\ref{fig:polarisation} shows the maximum polarisation,
as a function of wavelength, produced by an unresolved Earth-sized planet
orbiting at a distance of 0.01\,AU from a 5000\,K white dwarf.
We examined the dependence of the polarisation on orbital separation and
planet radius, taking the polarisation at 4000\,\AA. The results are shown
in the lower panels of Fig.~\ref{fig:polarisation}.

Table~\ref{tab:pol-values} gives the polarisation at three different
wavelengths, produced by Mars-sized, Earth-sized and Super-Earth-size planets
orbiting around a 5000\,K white dwarf. Three different distances from the
host star were used: 0.0054, 0.01, and 0.02\,AU, representative of
the inner edge, centre and outer edge of the white dwarf CHZ, respectively.
Table~\ref{tab:pol-values} also gives the polarisation of unresolved
Earth-sized planets orbiting in the habitable zone of a Sun-like star and
of a typical M dwarf (\Teff=3200\,K).

Table~\ref{tab:pol-values} shows that the polarisation produced by
an unresolved Earth-sized planet at 1\,AU from a Sun-like star is the order
of 10$^{-10}$. For a typical M dwarf, the habitable zone is at a distance of
$\sim$0.1\,AU \citep{selsis07}, therefore the amount of polarisation
is the order of 10$^{-8}$. The white dwarf CHZ is at 0.01\,AU,
consequently the amount of polarisation is the order of 10$^{-6}$. The
polarisation signal of any planet in the white dwarf CHZ
is therefore larger than that of a comparable planet in the habitable
zone of any other type of star, excluding brown dwarfs,
where low luminosity and high stellar activity most likely preclude life.
%
\begin{table}
\caption{\label{tab:pol-values} Polarisation obtained for a planet in the CHZ 
of a \Teff=5000\,K white dwarf. The polarisation is given as a function of 
planet radius (in \Rearth), orbital separation (in AU), and wavelength 
(in \AA). A planet of 0.5\,\Rearth\ would be about as large as Mars, while 
2.1\,\Rearth\ is the typical size of a Super-Earth. The orbital separations 
are the minimum and maximum limits, and center of the CHZ. For comparison, 
in the two bottom lines, we list the polarisation obtained for an Earth-size 
planet in the habitable zone at 1\,AU from the Sun (\Teff=5777\,K) and 
at 0.1\,AU from an M-dwarf (\Teff=3200\,K). For all calculations we assumed 
a phase angle of 68$^\circ$, where the polarisation is at maximum.}
\begin{footnotesize}
\begin{tabular}{cccc|ccc|ccc}
\tableline\tableline
 & \multicolumn{3}{c|}{Polarisation at 4000\,\AA} & \multicolumn{3}{c|}{Polarisation at 5000\,\AA} & \multicolumn{3}{c}{Polarisation at 6000\,\AA}\\
\tableline
 & \multicolumn{3}{c|}{Orbital separation in AU} & \multicolumn{3}{c|}{Orbital separation in AU} & \multicolumn{3}{c}{Orbital separation in AU}\\
\Rearth & 0.0054 & 0.01 & 0.02 & 0.0054 & 0.01 & 0.02 & 0.0054 & 0.01 & 0.02 \\
\tableline
0.5 & 6.6$\times$10$^{-7}$ & 1.9$\times$10$^{-7}$ & 4.8$\times$10$^{-8}$ & 3.2$\times$10$^{-7}$ & 9.3$\times$10$^{-8}$ & 2.3$\times$10$^{-8}$ & 1.4$\times$10$^{-7}$ & 4.1$\times$10$^{-8}$ & 1.0$\times$10$^{-8}$ \\
1.0 & 2.6$\times$10$^{-6}$ & 7.7$\times$10$^{-7}$ & 1.9$\times$10$^{-7}$ & 1.3$\times$10$^{-6}$ & 3.7$\times$10$^{-7}$ & 9.3$\times$10$^{-8}$ & 5.7$\times$10$^{-7}$ & 1.7$\times$10$^{-7}$ & 4.1$\times$10$^{-8}$ \\
2.1 & 1.2$\times$10$^{-5}$ & 3.4$\times$10$^{-6}$ & 8.4$\times$10$^{-7}$ & 5.6$\times$10$^{-6}$ & 1.6$\times$10$^{-6}$ & 4.1$\times$10$^{-7}$ & 2.5$\times$10$^{-6}$ & 7.3$\times$10$^{-7}$ & 1.8$\times$10$^{-7}$ \\
\tableline
1.0 & \multicolumn{3}{c|}{Sun at 1\,AU: 7.7$\times$10$^{-11}$} &  \multicolumn{3}{c|}{Sun at 1\,AU: 3.7$\times$10$^{-11}$} &  \multicolumn{3}{c}{Sun at 1\,AU: 1.7$\times$10$^{-11}$} \\ 
1.0 & \multicolumn{3}{c|}{M-dwarf at 0.1\,AU: 7.7$\times$10$^{-9}$}  &  \multicolumn{3}{c|}{M-dwarf at 0.1\,AU: 3.7$\times$10$^{-9}$}  &  \multicolumn{3}{c}{M-dwarf at 0.1\,AU: 1.7$\times$10$^{-9}$}  \\ 
\tableline
\end{tabular}
\end{footnotesize}
\end{table}
\section{Photometric Variations}
The planet will cause a variability of the total flux larger than that of 
the polarised flux. At phase angle $\alpha = 0$, (orbital phase
0.5, superior conjunction of the planet) the full disc of a planet at
$i=90^\circ$ will be illuminated producing a photometric maximum. The
photometric minimum occurs at inferior conjunction of the planet
($\alpha = 180^\circ$), with intermediate fluxes at quadrature
($\alpha = 90^\circ$), where the polarimetric maxima occur. The detailed
photometric light curve would depend on the planet scattering matrix,
but the reflected flux will be roughly proportional to the illuminated area.

Photometric variability can, however, arise from many different mechanisms, so
a planet detection based on photometry alone would not be credible. Since cool
non-magnetic white dwarfs are expected to have little intrinsic photometric
variability, a small photometric signal phased appropriately with the 
polarimetric variability could serve to confirm a polarimetric planet 
detection.

Planets in the CHZ of CWDs fill a substantial fraction of their Roche lobes.
We quantified this numerically, calculating the Roche geometry for planets
orbiting at 0.01\,AU from 0.6\,\M\ white dwarfs. Planets with masses and
radii similar to Earth, Corot-7b, Kepler-10b and Kepler-11b are all noticeably
distorted from spherical: the ratio of their illuminated area at maximum
and at quadrature is 1.95, 1.95, 1.97, and 1.90 respectively. A spherical
planet would have a ratio of exactly 2.0. A bloated planet similar
to Kepler-11d would overfill its Roche-lobe, and would thus lose its outer
layers until it exactly fills the critical equipotential surface. It would
present an enlarged teardrop-shaped cross-section at quadrature, with the
illuminated area at photometric maximum being only 1.52 times that at
quadrature.
\section{Reflected light searches for planets in the white dwarf CHZ}
\label{sec:advantage}
\citet{agol} made the case for transit searches targeting white dwarfs.
Planets in the CWD habitable zone would also reflect a relatively large
fraction of the stellar flux, producing also a relatively strong polarisation
signal, as quantified in Sect.~\ref{sec:pol}.
\clearpage
\begin{figure}
\includegraphics[width=16cm]{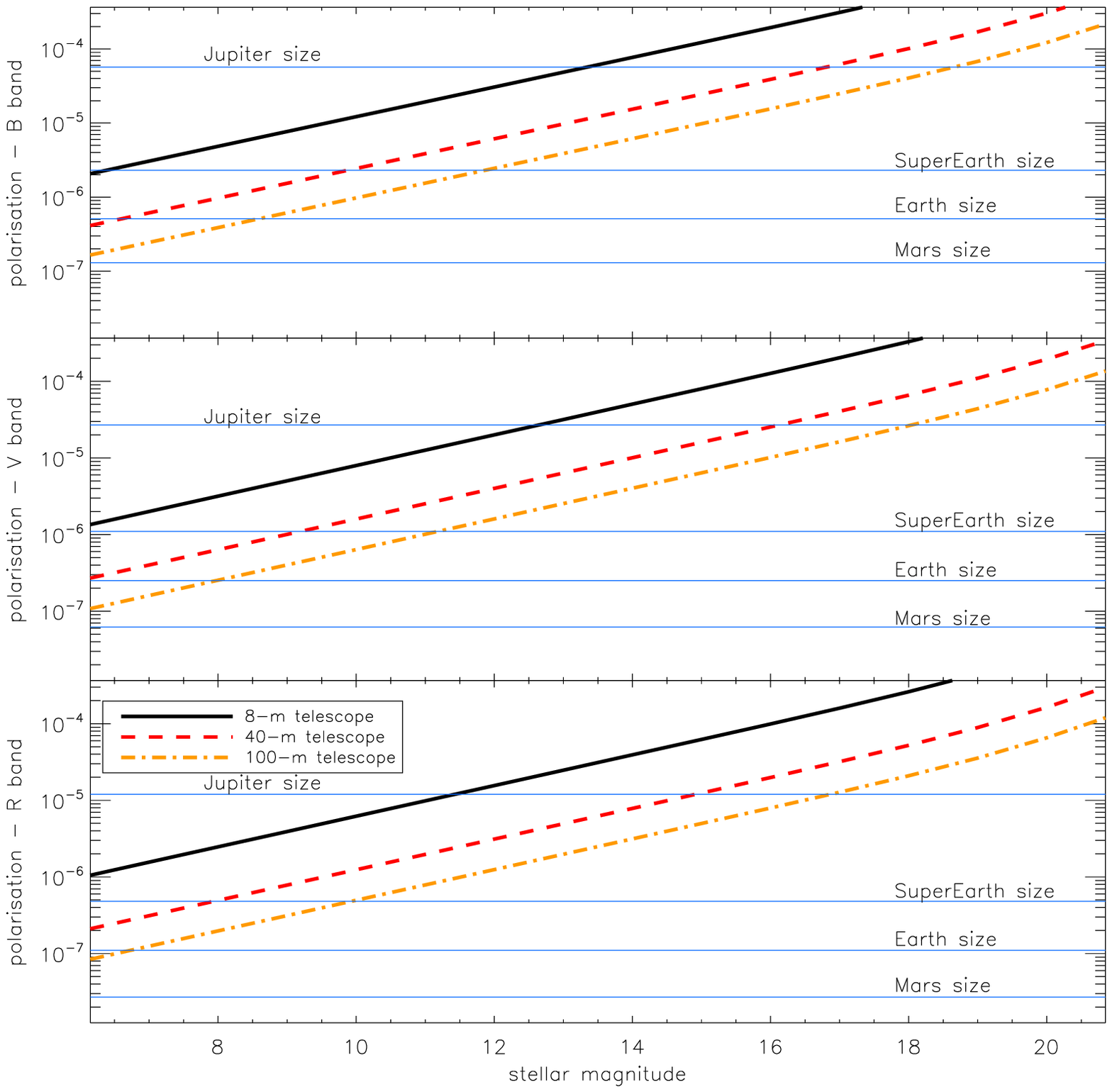}
\caption{\label{fig:VLT} Amount of polarisation detectable at 3$\sigma$ in
the $B$ (top), $V$ (middle) and $R$ (bottom panel) bands, as a
function of stellar magnitude, with a hypothetical FORS-like polarimeter 
mounted at a 8-m (black full line), 40-m (red dashed line), and 100-m (yellow
dash-dotted line) telescope, and an exposure time of 2.5\,hours. The horizontal
lines show the amount of polarisation emitted in the three bands by planets
of different sizes, orbiting at 0.01\,AU from a \Teff=5000\,K white dwarf.
Note: the amount of polarisation emitted by a Jupiter-sized planet has to be
taken with caution as we employed models by \citet{stam08}, which are not
designed for gas giants.}
\end{figure}

The CWD habitable-planet orbital separation, $a$, precludes planet detection
by direct imaging unless the distance, $d$, satisfies
\begin{equation}
\frac{d}{\rm pc} \leq 1.2 {\left ( \frac{a}{0.01 {\rm AU}}\right )} {\left ( \frac{400 {\rm nm}}{\lambda} \right )} {\left ( \frac{D}{10 {\rm m}}\right )}
\end{equation}
where $D$ is the aperture determining the diffraction limit and assuming the 
use of a coronograph working at an angular separation of $\frac{\lambda}{D}$. 
For the fiducial numbers above, even Proxima Centauri, the closest known star 
is too distant. Similarly, GAIA's astrometric precision could only detect an 
Earth-like planet in the CHZ for 0.6\,\M\ white dwarfs less than half the 
distance of Proxima Centauri. Microlensing too, is only sensitive to planets 
with wide orbits. For hydrogen and helium white dwarfs (the most common) the 
radial velocity detection of Earth-like planets in the white dwarf CHZ is 
precluded as the few available H and He lines do not facilitate the required 
precision of $<$1\,\ms. Reflected light (with ellipsoidal variation) and 
polarisation seem to be the most viable ways to detect close-in non-transiting 
rocky planets orbiting white dwarfs. 

Interstellar material could generate a much larger polarisation signal, but 
constant with time, hence easily disentanglable by that due to a planet. 
Magnetic fields and circumstellar discs could be responsible for small and 
time-dependent linear polarisation signals (although with a different 
wavelength dependency). Detection of polarimetric variability would not be 
sufficient to warranty planet detection. This should be supported by the 
combined analysis of the photometric and polarimetric curves as function of 
wavelength.

To detect linear polarisation at a 3$\sigma$ level in a rocky planet in the 
white dwarf CHZ we require observations at a signal-to-noise ratio (S/N) of 
$\sim$10$^6$.
White dwarfs are faint, constituting the largest obstacle to the planet
detection. Furthermore the very short planet orbital period ($\sim$12\,hours 
in the CHZ) limits integration times to 2--3\,hours.

Figure~\ref{fig:VLT} shows the polarisation detectable with a hypothetical
polarimeter (with transmission characteristics similar to those
of the FORS2 instrument at the ESO/VLT), mounted at a 8-m (e.g., VLT),
40-m (e.g., E-ELT) and 100-m (e.g., OWL) telescope, in the $B$, $V$,
and $R$ bands, as a function of stellar magnitude. We assumed a fixed
integration time of 2.5\,hours, a black body radiator of 5000\,K, new moon,
and an airmass of 1.6. As shown by \citet{berdyugina11}, the $B$ band compares
well with the $V$ and $R$ bands, because of the larger
polarisation emitted by the planet at shorter wavelengths, despite
lower flux and CCD sensitivity.

The brightest single CWD is WD\,0046+051 (\Teff=6220\,K), with a magnitude
of $V\sim$12.4 \citep{holberg08}. Current observing facilities are capable
of a $3\sigma$ detection only of a Jupiter-sized planet in the CHZ
of this star using an exposure time of 2.5\,hours. Future observing facilities
could detect Super-Earth-sized planets (between 2.0 and 2.5\,\Rearth), in the 
CHZ. Known white dwarfs with a temperature below 6000\,K are all of 
$V\geq$14\,mag and Fig.~\ref{fig:VLT} shows that for them even a 40-m 
telescope barely detects Super-Earth-sized planets in the CHZ, via polarimetry. 
This situation could be easily improved by using an instrument with a higher 
sensitivity in the blue.

For an atmosphereless planet, the amount of polarisation would be colour
independent \citep{bagnulo06}. Multiband polarimetry could then be used to
identify the presence or absence of a planetary atmosphere. For an
atmosphereless planet we could therefore integrate over the whole visible 
region with a considerable gain in S/N. This strategy will permit $3\sigma$
detection of Super-Earth-sized planets in the CHZ of a $V$=12\,mag CWD with
present-day observing facilities.

Summary: Non-magnetic white dwarfs have the characteristics necessary to
host the advent of life on planets orbiting in the continuous habitable zone.
A planet in the CHZ has 100 (10000) times the linear polarisation of the same
planet orbiting in the habitable zone of an M-dwarf (Sun-like star). The
relatively high polarisation of planets orbiting in the white dwarf CHZ and
the extreme stability of the linear polarisation emitted by non-magnetic
white dwarfs make polarimetry a potentially effective way to detect close-in
planets. In the case of atmosphere-bearing planets, the lack of bright white
dwarfs coupled with the short orbital period of planets in the CHZ,
only permit detections of Jupiter-sized planets with current observing
facilities (8-m telescope) and Super-Earth-sized planets with future facilities
(40-m and 100-m telescope). For atmosphereless planets, the observational
capabilities allow detection of Super-Earth-sized planets in the CHZ for the
brightest known cool white dwarf.

\end{document}